\documentclass[aps,preprint,noshowpacs,superscriptaddress,groupedaddress]{revtex4}  % for double-spaced preprint

\usepackage{graphpap}
\usepackage[dvips]{graphicx}
\usepackage[dvips]{graphics}
\usepackage{color}
\usepackage{hyperref}
\usepackage{amsmath}
\usepackage[normalem]{ulem}

\graphicspath{{./figures/}}
\usepackage{SIunits}
\usepackage{nomencl}
\makenomenclature

\begin{document}

%--------------------------------------------------
\title{Switching off energy decay channels in nanomechanical resonators}
%--------------------------------------------------

\author{J.~G\"uttinger}
\author{A.~Noury}
\author{P.~Weber}
\affiliation{ICFO-Institut de Ciencies Fotoniques, The Barcelona
Institute of Science and Technology, 08860 Castelldefels
(Barcelona), Spain}
\author{A.M.~Eriksson}
\affiliation{Department of Physics, Chalmers University of
Technology, S-41296 G{\"o}teborg, Sweden}
\author{C.~Lagoin}
\author{J.~Moser}
\affiliation{ICFO-Institut de Ciencies Fotoniques, The Barcelona Institute of Science and Technology, 08860 Castelldefels (Barcelona), Spain}
\author{C.~Eichler}
\author{A.~Wallraff}
\affiliation{Department of Physics, ETH Z\"urich, CH-8093 Z\"urich, Switzerland}
\author{A.~Isacsson}
\affiliation{Department of Physics, Chalmers University of
Technology, S-41296 G{\"o}teborg, Sweden}
\author{A.~Bachtold}
\thanks{Corresponding author: adrian.bachtold@icfo.es}
\affiliation{ICFO-Institut de Ciencies Fotoniques, The Barcelona Institute of Science and Technology, 08860 Castelldefels (Barcelona), Spain}

%\date{\today}

%\keywords{Optomechanics, graphene, mechanical resonator, NEMS, cavity readout}

\begin{abstract}
Energy decay plays a central role in a wide range of phenomena, such as optical emission, nuclear fission, and dissipation in quantum systems. Energy decay is usually described as a system leaking energy irreversibly into an environmental bath. Here, we report on energy decay measurements in nanomechanical systems based on multi-layer graphene that cannot be explained by the paradigm of a system directly coupled to a bath. As the energy of a vibrational mode freely decays, the rate of energy decay switches abruptly to a lower value. This finding can be explained by a model where the measured mode hybridizes with other modes of the resonator at high energy. Below a threshold energy, modes are decoupled, resulting in comparatively low decay rates and giant quality factors exceeding 1 million. Our work opens up new possibilities to manipulate vibrational states, engineer hybrid states with mechanical modes at completely different frequencies, and to study the collective motion of this highly tunable system.
\end{abstract}

\maketitle 
Energy decay is central in many fields of physics, including acoustics~\cite{Rayleigh1871}, non-equilibrium thermodynamics~\cite{Onsager1931}, and
the quantum mechanics of dissipative systems~\cite{Feynman1963,Caldeira1983}. A
dissipative system is always coupled to a given thermal environmental
bath. The rate of the coupling is often considered to be independent of the system energy. Recently, nonlinear phenomena associated with dissipation have attracted considerable interest. These include measurements on radio-frequency superconducting resonators, where the energy-dependent decay rate is attributed to the saturation of two-level defect states ~\cite{Gao2007,Burnett2014}. In addition, measurements on nanomechanical resonators have been described by a phenomenological nonlinear decay process~\cite{Eichler2011, Zaitsev2012, Imboden2013, Mahboob2015, Polunin2016}. However, the physical mechanism responsible for the nonlinear dissipation remains elusive. Understanding the origin of dissipation in mechanical resonators is crucial ~\cite{Dykman1984,cleland2002,Wilson-Rae2008,Rieger2014,Tao2014,Midtvedt2014}, because it is a key figure of merit for many applications, such as mass, force, and spin sensing~\cite{GilSantos2010, Hanay2012,chaste2012,Rugar2004,Moser2014}. Here, we report energy decay measurements on mechanical resonators that are radically different from what is usually observed, and that allow us to reveal the nature of the nonlinear energy decay process.  

\begin{figure*}[t]
    \includegraphics[width=0.9\linewidth]{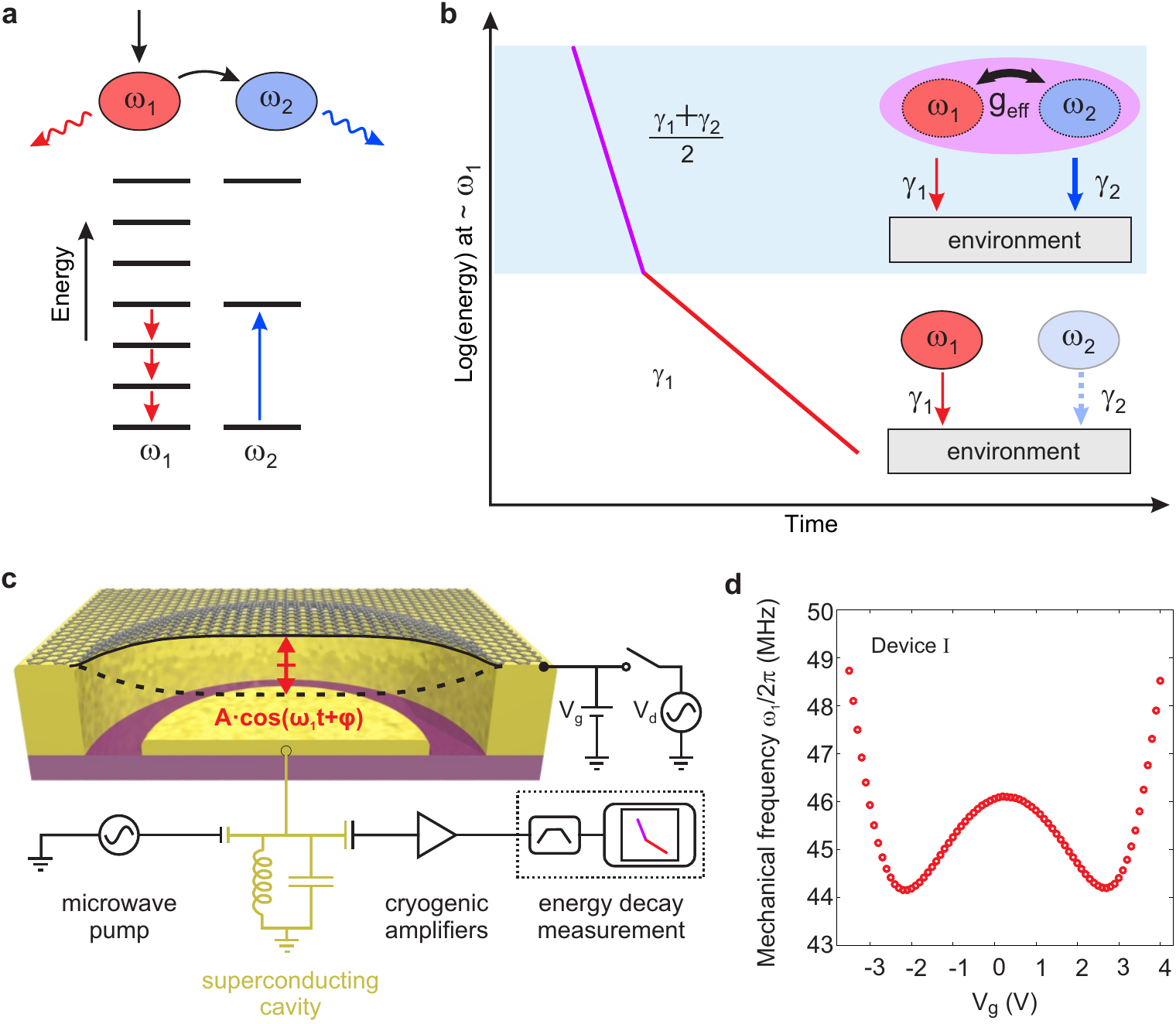}%0.9
    \caption{\textbf{Nonlinear energy decay process, mode hybridization, and graphene resonator}. 
    \textbf{a}, Energy diagram showing an energy exchange process between two harmonic oscillators. In the top of the figure, we depict the flow of energy when driving mode 1 strongly.
    \textbf{b}, Energy decay measured at the frequency of mode~1. Because of the nonlinearity, the effective coupling strength depends on energy. At high energy, the modes are hybridized and decay in unison with rate $(\gamma_1+\gamma_2)/2$. As energy decays, the modes decouple, resulting in a change of the decay rate for mode 1. \textbf{c}, Measurement setup with schematic cross-section of a circular graphene drum vibrating as $\cos(\omega_\mathrm{1} t + \varphi)$ where $\varphi$ is the phase relative to the capacitive driving force. The motion is detected with the superconducting microwave cavity. After cryogenic amplification, the output signal of the cavity is recorded and digitally computed to obtain the energy and the frequency of the vibrations as a function of time. The application of the static voltage $V_\mathrm{g}$ between the graphene drum and the cavity electrode allows us to tune the mechanical resonance frequencies; the oscillating voltage $V_\mathrm{d}$ is used to drive the resonator. 
    \textbf{d}, Frequency of the fundamental mode as a function of gate voltage $V_{\rm g}$.}
    \label{figIntro}
\end{figure*}

To gain insight into this nonlinear process, we consider the way by which energy stored in a vibrating guitar string is dissipated. In essence, the sound coming from the guitar originates from the coupling of the vibrating string with the air in the room, which acts as the environmental bath. To increase the intensity of the sound, and to quickly tone the sound down, the string is coupled resonantly to the guitar body. As the string and the body vibrate at similar frequencies, the vibrational energy is transferred efficiently. The coupling is said to be linear, in a sense that the coupling energy $H_c$ is proportional to the product of the vibration amplitude of the string $q_1$ by the vibration amplitude $q_2$ of the guitar body, $H_c\propto q_1q_2$. Nonlinear couplings can also exist, and have the peculiarity of enabling energy transfer between vibrational modes even if resonant frequencies are far apart. For this to happen, the ratio of resonant frequencies $\omega_2/\omega_1$ has to be close to an integer $n$~\cite{Antonio2012,Eichler2012,nayfeh2008nonlinear}. Although the modes are in the classical regime, this nonlinear mode coupling can be easily understood by considering the energy ladders of harmonic oscillators. Figure~\ref{figIntro}a illustrates an energy conserving process (here $n=3$) annihilating simultaneously three quanta in mode 1, while creating one quantum in mode 2. The lowest order nonlinear coupling term in the Hamiltonian that can achieve this exchange is $H_\mathrm{c}\propto q_1^3q_{2}$. Nonlinear coupling is particular, as the coupling strength is energy dependent. At high vibrational amplitude, the modes hybridize, and decay in unison with rate $(\gamma_1+\gamma_2)/2$. At low amplitude, the two modes are decoupled and decay with rate $\gamma_1$ and $\gamma_2$, respectively (Fig.~\ref{figIntro}b). As a result, the flow of energy takes different paths during an energy decay measurement of mode 1. At high amplitude, the energy of the vibrations is transferred into the bath of mode~1 directly, and into the vibrations and the bath of mode~2. At low amplitude, the dissipation channel to the bath of mode~2 is closed. This abrupt variation of the mechanical decay has to our knowledge neither been predicted nor been measured. The associated nonlinear energy decay process is generic and should be relevant to a large variety of systems, such as mechanical, optical, and electrical resonators.

\begin{figure}
    \includegraphics[width=.5\linewidth]{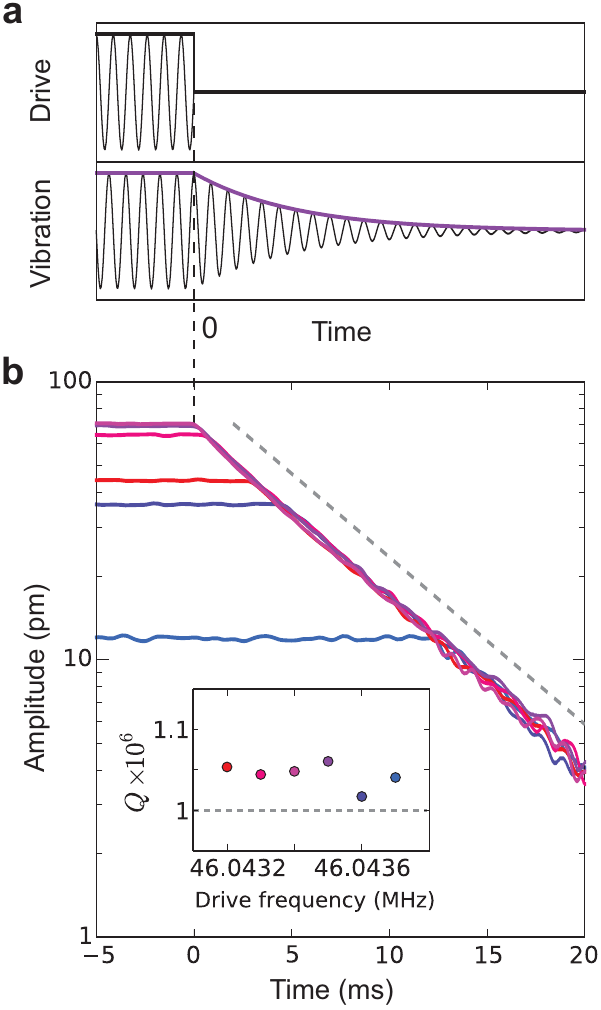}%0.5
    \caption{\textbf{Energy decay measurements of a graphene resonator with a $Q$-factor of 1 million in the low vibrational amplitude regime.}
    \textbf{a}, Measurement principle. At time $t=0$ the mechanical driving force is switched off and the vibrational amplitude starts to decay.
    \textbf{b}, Measured energy decay of the vibrational amplitude of device I as a function of time for different drive frequencies (see colors in inset). The graphene membrane is 5-6 layer thick. The lower amplitude traces are shifted in time so all decaying curves overlap. The dashed grey line indicates an exponential decay corresponding to a $Q$-factor of 1 million. The inset shows the quality factor as a function of drive frequency. We apply $V_\mathrm{g}=
    0.6~$V and pump the cavity with $n_\mathrm{p} \approx 1000$ photons.\label{FigHighQ}}
\end{figure}

\textbf{Measuring the energy decay of graphene resonators.} Graphene-based resonators are well suited for observing such a nonlinear energy decay process (Figs.~\ref{figIntro}c,d). Indeed, their resonant frequencies can be widely tuned by electrostatic means~\cite{Eichler2011,Chen2009,Singh2010,Barton2012,Miao2014,DeAlba2016,Mathew2016} allowing us to set the frequency ratio of two modes to an integer. Energy decay measurements are carried out by preparing the resonator in an out-of-equilibrium state using a capacitive driving
force, then switching the drive off and measuring the vibrational
amplitude as the mechanical energy freely decays
(Fig.~\ref{FigHighQ}a). A circular graphene
mechanical resonator is capacitively coupled to a superconducting microwave cavity
in order to detect the mechanical
vibrations with a short time resolution, a high displacement
sensitivity, and over a broad range of vibrational
amplitudes~\cite{Weber2014, Song2014, Singh2014,Weber2016}. The time
resolution is limited by the inverse of the coupling rate of the
cavity to the external readout circuit, which is of the order of
$\kappa_\mathrm{ext}/2\pi \approx 1$~MHz in our devices. High
displacement sensitivity with minimal heating is demonstrated by
resolving thermal motion at about 50~mK, corresponding to $\approx
25$ quanta of vibrations. The displacement
sensitivity can be further improved using a near quantum-limited
Josephson parametric amplifier (JPA) for the readout of devices
with comparatively low signal output~\cite{Eichler2014}. We typically record vibrations with amplitudes ranging from 1~pm to 1~nm. Using a time-resolved
acquisition scheme combined with real-time digital signal
processing (dashed box in Fig.~\ref{figIntro}c), we
record the two quadratures of motion, which are digitally squared
to compute the vibrational energy. Energy decay traces are
obtained by averaging typically 1000 measurements and subtracting
a time independent noise background that is related to the
amplifier chain. Decay traces are displayed as vibration amplitude versus time to make contact with other experiments. All the data presented here are taken at 15~mK. See Supplementary Information (SI), Secs.~1-5, for further details on the devices, the measurement setup and the displacement calibration.

\textbf{Quality factor above 1 million.}
In the low-amplitude regime, this measurement scheme is beneficial as it allows us to observe record-high quality factors (Fig.~\ref{FigHighQ}b). The measured vibration amplitude of the fundamental mode at frequency $\omega_1$ 
decays exponentially in time as $ \propto e^{-
\gamma_1t/2 }$ with an energy decay rate
$\gamma_1 \approx 1/(3.6~\textrm{ms})$. This corresponds to a quality factor $Q$ exceeding 1
million, surpassing previously reported $Q$-factors in graphene~\cite{Eichler2011, Weber2014,Singh2014}. 
By collecting energy decay traces using different drive frequencies near $\omega_\mathrm{1}$, we show that the $Q$-factor is independent of the drive frequency and the
vibrational amplitude at the beginning of the ring-down
(Fig.~\ref{FigHighQ}b). One reason for the observation of such
high $Q$-factors is the fact that our technique is immune from
dephasing, in contrast to previous spectral measurements on
graphene resonators. Comparing energy decay measurements with
spectral thermal motion measurements reveals that dephasing is
significant, as the resonance linewidth is more than twice as
large as the energy decay rate (Sec.~6 of SI).

\begin{figure}
    \includegraphics[width=\linewidth]{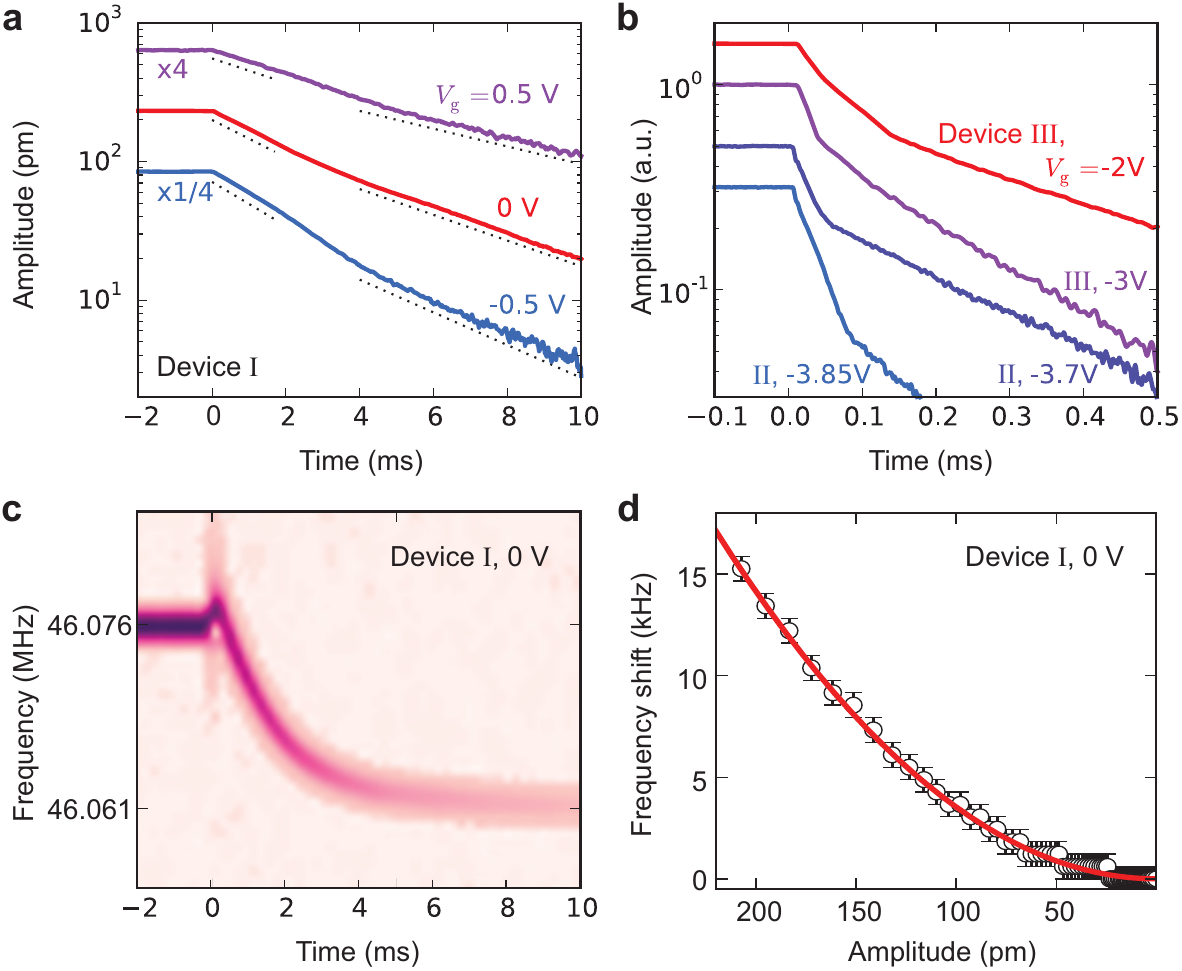}
    \caption{\textbf {Energy decay in the high vibrational amplitude regime.}
        \textbf{a},\textbf{b}, Energy decay measurements at different $V_\mathrm{g}$ for three different devices (labeled I, II, and III). In all cases, the decay rate changes discontinuously. The bandwidth of the bandpass filter in \textbf{a} is $150~$kHz for the violet and red traces and $200~$kHz for the blue trace. The bandwidth is $400~$kHz and $200~$kHz for device II and III in \textbf{b}, respectively. \textbf{c}, Time dependence of the short-time Fourier transform of the vibrations corresponding to the red amplitude decay trace in \textbf{a}. \textbf{d},  Frequency shift as a function of vibrational amplitude. The quadratic dependence (red line) is in agreement with the frequency pulling expected from the nonlinear restoring force at low vibration amplitude.
        \label{FigKink}}
\end{figure}

\textbf{Nonlinear behaviour in the energy decay.}
In the high-amplitude regime, our high-precision
measurements reveal a discontinuity in the energy decay rate
during the ring-down (Figs.~\ref{FigKink}a,b). This finding is robust, since it is
observed in all the studied resonators. Changing the amplitude and frequency of the initial driving force does not affect the rates and amplitudes associated with the discontinuities. However, what strongly affects the energy-decay traces is the static voltage $V_\mathrm{g}$ applied between the graphene
resonator and the cavity. This is a first indication that the discontinuity of
the decay is related to nonlinear mode coupling, since the variation
of $V_\mathrm{g}$ strongly modifies the different resonant frequencies. To ensure that all the vibrational energy during the decay is properly captured, we use signal filtering with large bandwidth. This is because the frequency changes during the decay (Figure~\ref{FigKink}c). Upon comparing this smooth frequency change and the vibrational amplitude decay, we get the expected quadratic amplitude dependence of the frequency related to the nonlinear Duffing restoring force~\cite{lif08} (Fig.~\ref{FigKink}d and Sec.~7 of SI).

\begin{figure*}[t]
    \includegraphics[width=0.9\linewidth]{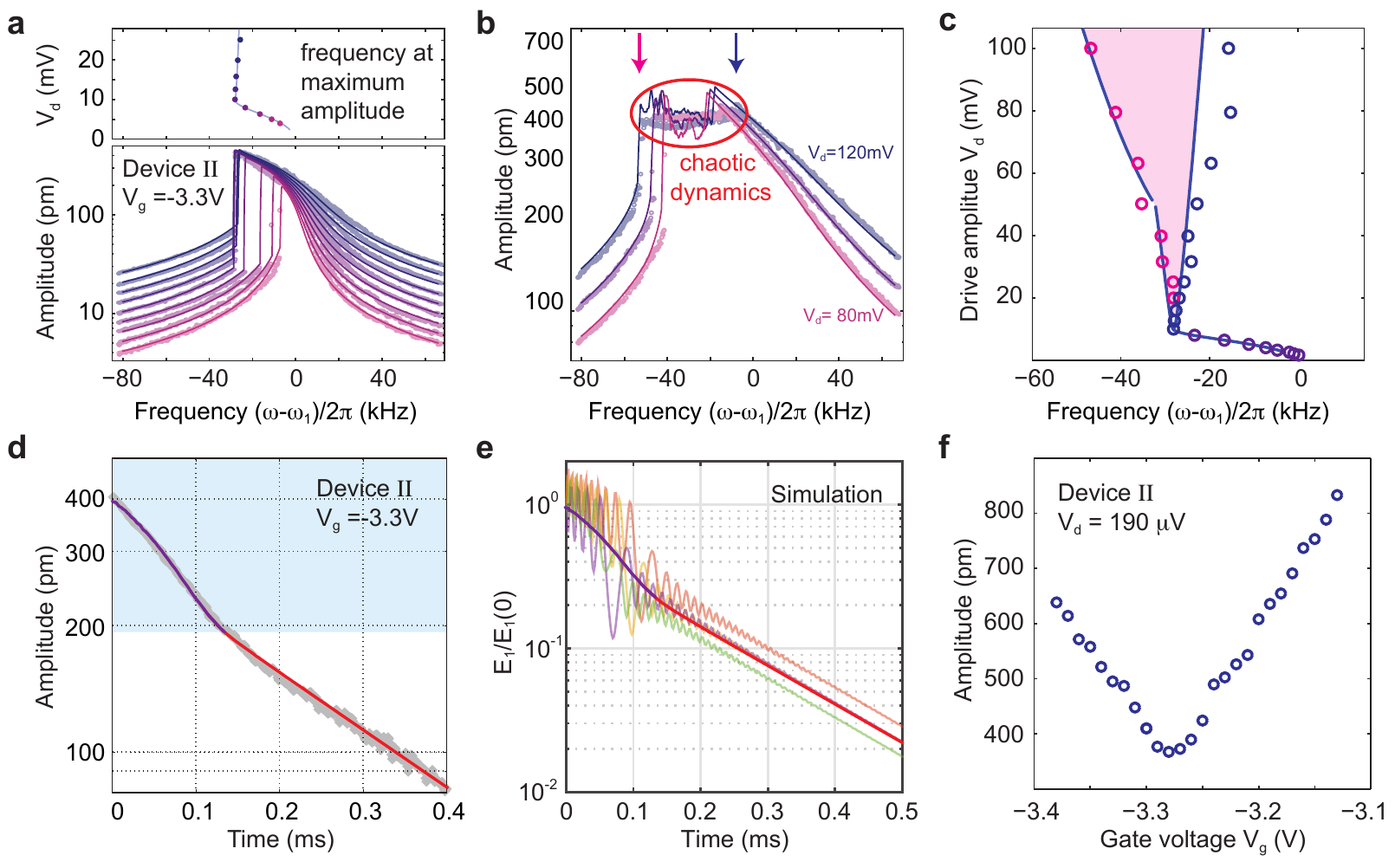}
    \caption{\textbf {Driven response and energy decay traces.}
        \textbf{a}, Measured and simulated driven response of the fundamental mode for intermediate drive strengths (lower panel), the corresponding measured and calculated bifurcations where the high-amplitude states switch to the low-amplitude states (upper panel). The drive is swept from high frequency to low frequency. The observed saturation of the switching frequency at high drive occurs when $\omega_2=3\omega_1$, providing strong evidence for mode interaction.
        \textbf{b}, Measured and simulated driven response of the fundamental mode for strong driving. The plateau is associated with chaotic dynamics of coupled nonlinear modes. The blue and pink arrows indicate the location of the experimentally detectable bifurcations in this regime with a downward frequency sweep.
        \textbf{c}, Reduced bifurcation diagram showing the experimentally accessible bifurcations (measured and calculated) for a downward frequency sweep. The data points are obtained from driven response measurements at different drive strengths; see \textbf{a} and \textbf{b}. A more complete diagram can be found in Sec.~8.1 of the SI.
        \textbf{d}, Measured and simulated energy decay traces. Using the same parameters as in the fits in \textbf{a}-\textbf{c}, the model satisfactorily reproduces the measured energy decay trace, including the discontinuity in the decay rate.
        \textbf{e}, Simulated individual energy decay traces revealing coherent oscillations. Different traces correspond to different initial states, which are all prepared with the same drive strength. The calculated trace in \textbf{d} is obtained from averaging over many ring-downs. 
        \textbf{f}, Vibration amplitude of the plateau feature in \textbf{b} as a function of gate voltage $V_{\rm g}$. The plateau is a consequence of mode hybridisation (Sec.~8.1 of SI). The amplitude is extracted from the measured response at the frequency just below the bifurcation indicated by the blue arrow in \textbf{b}.
        \label{hybridization}}
\end{figure*}

We characterize mode coupling by measuring the response of the fundamental mode to the driving force. For this, we tune $V_{\rm g}$ so the frequency $\omega_1$ of the fundamental mode is about one third the frequency $\omega_2$ of a higher-order mode, that is, $\omega_1/2\pi=\unit{44.132}{\mega\hertz}$ and  $\omega_2/2\pi=$\unit{132.25}{\mega\hertz}. For intermediate drive amplitudes $V_{\rm d}$, the response of mode 1 is that of a Duffing resonator with a softening nonlinearity
(Fig.~\ref{hybridization}a). However, as driving increases, we observe a saturation of the frequency at which the high-amplitude state switches to the low-amplitude state. The related bifurcation points are shown in the top part of Fig.~\ref{hybridization}a. This saturation of the frequency is due to the efficient energy transfer between mode~1 and mode~2 when the frequency ratio is an integer~\cite{Antonio2012}. Driving the system even harder, the response exhibits a plateau behaviour as shown in Fig.~\ref{hybridization}b.  

We now show that the abrupt variation of the decay rate is related to nonlinear mode coupling. To demonstrate this, we use the driven response measurements shown above to determine the parameters of a minimal model of coupled nonlinear resonators, which allows us to describe the measured energy decay with good accuracy, as discussed next. Both features, frequency saturation and plateau, are well reproduced by the minimal model of two coupled nonlinear modes 
\begin{eqnarray}
&&\ddot{q}_1=-\omega_1^2q_1-\gamma_1\dot{q}_1-\alpha_1q_1^3-m^{-1}\partial H_\mathrm{c}/\partial q_1+aV_\mathrm{d},\label{eq:q1}\\
&&\ddot{q}_2=-\omega_2^2q_2-\gamma_2\dot{q}_2-\alpha_2q_2^3-m^{-1}\partial H_\mathrm{c}/\partial q_2,\label{eq:q2}
\end{eqnarray}
with $\alpha_1$ and $\alpha_2$ the Duffing constants, and $a$ the force constant. The interaction Hamiltonian is $H_{\rm c}=mgq_1^3q_2$ with $m\approx 60$~fg the effective mass of mode 1. The displacement $q_2$ of mode 2 is normalized to have the same effective mass. In these equations we only include terms necessary to capture the essential physics. We thus omit higher-order restoring force terms, off-resonant interaction terms, and purely dispersive coupling terms. Although the omitted terms may give a better fit, we seek for clarity to minimize the number of free parameters. As can be seen from Figs.~\ref{hybridization}a-c, the model allows us to describe the measurements of the response and the reduced bifurcation diagram with good agreement (Sec.~8 of SI). Remarkably, the parameters used to fit these driven measurements reproduce quantitatively the measured energy decay trace (Fig.~\ref{hybridization}d). This shows that nonlinear mode coupling is at the origin of the observed discontinuity in the decay rate. Note that single energy decay traces are expected to feature oscillations because of mode coupling (Fig.~\ref{hybridization}e). But our simulations show that these oscillations disappear when averaging multiple traces, in agreement with our experiments. For the strong drives used for the excitation of the resonator, mechanical nonlinearities cause the initial state to be different from  one decay measurement to the next, so that the oscillations in the averaged decay trace are washed out.

The discontinuity in the decay rate is related to the crossover from mode hybridization to weak mode coupling as the vibration amplitude freely decays. Indeed, the measured crossover amplitude of $\sim\unit{200}{\pico\meter}$ is consistent with the $\sim\unit{300}{\pico\meter}$ obtained from the crude estimate $g_\mathrm{eff}^2\sim (\gamma_1-\gamma_2)^2/4+\delta_\mathrm{eff}^2$. Here, $g_{\rm eff}=gm\omega_1^2A_1^2/2$ is the rate of energy exchange between the two modes with $A_1$ the amplitude of mode 1; and $\delta_{\rm eff}$ is the offset of frequency between the two modes including Duffing shifts (see Sec.~8.4 of SI).

The coupling constant $g$ obtained from our simulations is in reasonable agreement with the value expected from continuum elasticity. In the case of a circular membrane clamped at the circumference, the strength of the fourth order coupling expected from continuum elasticity yields $mg\sim NE_\mathrm{2d}/R^2$ with a prefactor that depends on the shape of the modes~\cite{Eriksson2013}. Using $N=35$ as the layer number of the graphene flake in device II, $E_\mathrm{2d}=340\,$Nm$^{-1}$ the 2-dimensional Young modulus of graphene, $R=1.6\,\mu$m the radius of the resonator, and $mg=7\cdot10^{16}\,$Jm$^{-4}$ the coupling obtained from our simulations, we get a reasonable prefactor of $14$~\cite{Eriksson2013}. The nonlinear constant $\alpha_1$ deduced from our simulations is consistent with the value of a graphene flake that is slightly bended by the static capacitive force~\cite{Weber2014}. All the parameters of the coupled system can be found in Sec.~8.2 of the SI.

The mode frequency ratio does not need to be strictly an integer to observe discontinuity in the decay rate. The discontinuity is indeed measured when detuning $V_{\rm g}$ so that $\omega_2\neq3\omega_1$ (Sec.~7 of SI). However, the hybridization occurs at larger amplitudes (Fig.~\ref{hybridization}f). Energy decay traces with discontinuity can be measured in very different $V_{\rm g}$ ranges, indicating that the frequency ratio of the mode coupling is close to an integer that can be different from three.

\textbf{Conclusions.}
In summary, we show that the oscillator-bath coupling paradigm employed to describe energy decay in mechanical resonators breaks down when modes at different frequencies hybridize. This effect is not a curiosity limited to a restrained parameter space, since the effect of hybridization is observed here for vibration amplitudes as low as $\sim$ 100~pm, while graphene and nanotube resonators are driven to the $1-10$~nm amplitude range in many experiments. It will be intriguing to investigate hybridization driven by thermal noise. For this, nanotube resonators are appealing~\cite{Moser2014,Benyamini2014}, since the variance of thermal vibrations can be well above $(1~\textmd{nm})^2$ at room temperature. It will be exciting to hybridize three and more modes by setting the appropriate gate voltage or by strongly driving the resonator. The collective motion will feature rich physics with the interplay of hybridization, Duffing nonlinearity, and noise. In addition, hybridization opens up new possibilities to tune dissipation by electrostatic means and to manipulate vibrational states coherently~\cite{DeAlba2016,Mathew2016,Faust2013,Okamoto2013,Xu2016}.

When preparing the manuscript, we learned about another work on the coupled-mode system described by Eqs.~(\ref{eq:q1}) and~(\ref{eq:q2}). The measurements reveal a mechanical frequency comb, a result completely different from ours, showing the richness of `internal resonances'~\cite{Shoshani2016}.

\textbf{Acknowledgments} We thank M. Dykman, S. Shaw, D. Lopez, F. Guinea, and N. Noury for discussions. We acknowledge Gustavo Ceballos and the ICFO mechanical and electronic workshop for support. We acknowledge financial supports by the ERC starting grant 279278 (CarbonNEMS), the EE Graphene Flagship (contract no. 604391), from MINECO and the ‘Fondo Europeo de Desarrollo Regional’ (FEDER) through grant MAT2012-31338, and the Generalitat through AGAUR. AI and AME acknowledge financial support through the Swedish research council (VR).

{\textbf{Contributions}
PW fabricated the devices. JG, AN, and PW carried out the experiment with support from CL and JM. Theoretical modeling and simulations were done by AME and AI. The JPA was provided by CE and AW. The data analysis was done by JG, AN, PW, AME, AI and AB. JG, AI and AB wrote the manuscript with comments from the other authors. AB supervised the work.}

%\section*{References}
%\bibliographystyle{Nature}

\end{document}